\documentclass[journal]{IEEEtran}

\usepackage[T1]{fontenc}

\usepackage[hidelinks]{hyperref}

\ifCLASSINFOpdf
  \usepackage[pdftex]{graphicx}
  \graphicspath{{./figures/}}
  \DeclareGraphicsExtensions{.pdf,.png,.jpeg}
\else
  \usepackage[dvips]{graphicx}
  \graphicspath{{./figures/}}
  \DeclareGraphicsExtensions{.eps}
\fi

\usepackage[cmex10]{amsmath}
\usepackage{array}
\usepackage{url}

\usepackage[absolute]{textpos}

\begin{document}

\begin{textblock}{15}(0.5,15.0)
{
\noindent\hrulefill

\noindent\fontsize{8pt}{8pt}\selectfont\copyright\ 2015 IEEE. Personal use of this material is permitted. Permission from IEEE must be obtained for all other uses, in any current or future media, including reprinting/republishing this material for advertising or promotional purposes, creating new collective works, for resale or redistribution to servers or lists, or reuse of any copyrighted component of this work in other works. \hspace{5pt} This is the accepted version of: M. Sul\'ir, J. Porub\"an. Trend analysis on the metadata of program comprehension papers. 13th International Conference on Engineering of Modern Electric Systems (EMES), IEEE, 2015, pp. 153--156. \url{http://doi.org/10.1109/EMES.2015.7158425}

}
\end{textblock}

\title{Trend Analysis on the Metadata\\of Program Comprehension Papers}
\author{Mat\'u\v{s}~Sul\'ir and~Jaroslav~Porub\"an
\thanks{Manuscript received March 30, 2015; revised May 12, 2015. This work was supported by VEGA Grant No.~1/0341/13 Principles and methods of automated abstraction of computer languages and software development based on the semantic enrichment caused by communication.}
\thanks{Mat\'u\v{s}~Sul\'ir (matus.sulir@tuke.sk) and Jaroslav~Porub\"an (jaroslav.poruban@tuke.sk) are with the Department of Computers and Informatics, Faculty of Electrical Engineering and Informatics, Technical University of Ko\v{s}ice.}}

\markright{}


\vspace{-100pt}
\maketitle
\pagenumbering{gobble}

\begin{abstract}
As program comprehension is a vast research area, it is necessary to get an overview of its rising and falling trends. We performed an n-gram frequency analysis on titles, abstracts and keywords of 1885 articles about program comprehension from the years 2000--2014. According to this analysis, the most rising trends are feature location and open source systems, the most falling ones are program slicing and legacy systems.
\end{abstract}


\section{Introduction}

\IEEEPARstart{P}{rogram} comprehension deals with an understanding of existing programs by developers. It is a vast research area, ranging from studying mental models to the design and evaluation of reverse engineering tools. For this reason, it is necessary to gain an overview of the field -- its methods and techniques. It is important to become familiar not only with stable knowledge, but also with the latest trends.

We have the following research questions:

\begin{itemize}
\item \textbf{RQ1:} What are the most rapidly rising trends in program comprehension?
\item \textbf{RQ2:} What are the falling trends in program comprehension?
\end{itemize}

\section{Method}

\begin{figure}[h]
\centering
\includegraphics[]{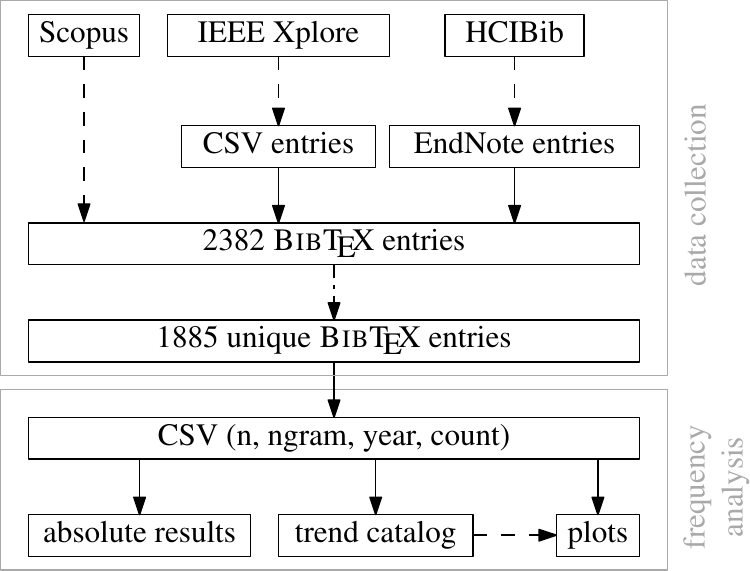}
\caption{A block diagram of the analysis method. Solid arrows represent automated processes, dash-dotted semiautomatic and dashed manual ones.}
\label{fig:process}
\end{figure}

A brief overview of the process is depicted in Fig.~\ref{fig:process}.

\subsection{Data Collection}

First, we searched the citation databases Scopus and IEEE Xplore for the following exact terms (in quotes):
\begin{itemize}
\item program comprehension,
\item program understanding,
\item code comprehension.
\end{itemize}
Only articles published between the years 2000 and 2014 (inclusive) were searched. The Scopus queries were continuously refined to exclude unrelated entries, e.g. articles about TV program comprehension or a chemical paper where the words ``program'' and ``understanding'' were only accidentally subsequent. In Scopus, excluding artifacts like proceedings cover pages, tables of contents and author indexes was easily accomplished by limiting the document types to journal articles and conference papers. For IEEE Xplore, an extensive query based on string matching had to be constructed.

Neither Scopus nor IEEE Xplore contained metadata from the last year of IEEE International Conference on Program Comprehension (ICPC 2014) which could negatively affect the validity of our study. They were available only in the ACM Digital Library, which does not offer an option to mass-download metadata and even prohibits it. Fortunately, HCI Bibliography, hosted by ACM SIGCHI, offered them\footnote{\url{http://hcibib.org/ICPC14}}.

We downloaded 1599 entries from Scopus in the \textsc{Bib}\TeX{} format, 740 from IEEE Xplore as a CSV (comma-separated values) file and 43 from the HCI Bibliography website in the EndNote format. We did not utilize IEEE Xplore's \textsc{Bib}\TeX{} export as it is limited to 100 entries and does not distinguish between author-supplied and automatically assigned keywords. All exported files were converted if needed and combined into one \textsc{Bib}\TeX{} file. Ten incomplete items (without authors or an abstract) and 487 duplicates were removed. This gives us a total of 1885 analyzed articles.


\subsection{Conversion to an N-gram List}

Each article in a bibliographic database is called an \textit{entry}. Every entry has \textit{fields} like an article title or year published.

For the purpose of this study, we analyzed titles, abstracts and author-supplied keywords using a technique called n-grams. An n-gram with a length $n$ is a sequence of $n$ words in a text. For example, the sentence ``Here you are'' contains \textit{unigrams} ``here'', ``you'' and ``are''; \textit{bigrams} ``here you'' and ``you are'' and a \textit{trigram} ``here you are''.

First, each abstract was split into a list of sentences to prevent n-gram analysis across sentence boundaries like recognizing the bigram ``program comprehension'' in the fragment ``to program. Comprehension of''. Although titles usually contain only one sentence, they were split, too. In our study, every keyword was considered a separate sentence.

Next, the words ``a'', ``an'' and ``the'' were removed from all sentences because the meaning of an n-gram like ``the program'' is the same as ``a program''. A list of n-grams was produced for each sentence. We worked with n-grams with a length from 1 to 4 as longer n-grams are rarely repeated in texts.

From the list of n-grams, we removed ones which comprised of at least 50\% stopwords. A \textit{stopword} is a frequent function word, i.e. a word which does not convey any meaning on its own. For example, the n-gram ``any of programs'' was excluded while the phrase ``comprehension of programs'' was retained. We used a stopword list from the Snowball stemmer\footnote{\url{http://snowball.tartarus.org/algorithms/english/stop.txt}}.

All n-grams with their associated origin information were transformed into a simple tabular model (a list of records) and saved into a CSV file. Each of its records contains these fields:
\begin{itemize}
\item n,
\item n-gram,
\item year,
\item count.
\end{itemize}

For example, a record ``2,dynamic analysis,2008,33'' means that a bigram ``dynamic analysis'' occurred 33 times in a title, abstract or keywords of articles published in 2008.

Up to this point, we used the Ruby language (mainly because there is a convenient BibTeX library available\footnote{http://rubygems.org/gems/bibtex-ruby}) to automatize some parts of the process.

\subsection{Frequency Analysis}

The produced CSV file was loaded by a script written in the statistical language R, which also generated the plots used in this article.

We defined a \textit{frequency} of an n-gram of length $n$ as follows:
\[
freq(ngram, year) = \frac{count(ngram, length(ngram), year)}{count(*, length(ngram), year)}
\]
where $count$ is a sum of the counts for records matching the given criteria and $*$ means ``any n-gram''. For example, the frequency of a bigram ``design pattern'' for year 2001 is the count of occurrences of this n-gram in 2001 divided by the sum of occurrences of all bigrams in this year. This approach is similar to the one used by the current version of Google Ngram Viewer \footnote{\url{http://books.google.com/ngrams/info}}.

Using this metric, a PDF file which we call \textit{Trend Catalog} was produced, containing year-frequency plots of about 800 most frequent n-grams in the whole database. We manually inspected the catalog to get an overview of the most interesting trends.

It is often useful find out the frequency of multiple closely related n-grams, for example {\small\texttt{slice+slices+slicing}}. Thus we define the frequency of a list of n-grams as the sum of the frequencies of individual n-grams:
\[
frequency(ngrams, year) = \sum_{x \in ngrams} freq(x, year)
\]

Finally, it is desirable to put multiple competing n-grams (or n-gram lists) to one plot. Competing phrases are separated by commas. We can perceive this as a simple query language, similar to the one used by Google Ngram Viewer \cite{Michel11quantitative}. For example, a plot of the query {\small\texttt{case study, experiment, review+survey}} is shown in Fig.~\ref{fig:methods}.

\section{Absolute Results}

Before exploring the actual trends, let us outline the results in terms of the absolute numbers of occurrences of the most frequent n-grams. The top three unigrams were ``program'', ``software'' and ``code''.

The most interesting were bigrams, shown in Table~\ref{tab:absolute}. High positions of the phrases ``reverse engineering'' and ``software maintenance'' correspond with our previous statement that they are two most related research fields to program comprehension \cite{Sulir15program}.

\begin{table}[b]
\renewcommand{\arraystretch}{1.3}
\caption{The Most Frequent Bigrams}
\label{tab:absolute}
\centering
\begin{tabular}{|r|l|r|} \hline
\textbf{Order} & \multicolumn{1}{c|}{\textbf{Bigram}} & \textbf{Count} \\ \hline
1. & program comprehension & 1541 \\ \hline
2. & source code & 1070 \\ \hline
3. & program understanding & 642 \\ \hline
4. & reverse engineering & 518 \\ \hline
5. & software maintenance & 357 \\ \hline
6. & software systems & 345 \\ \hline
7. & software system & 265 \\ \hline
8. & software engineering & 234 \\ \hline
9. & dynamic analysis & 208 \\ \hline
10. & case study & 191 \\ \hline
11. & program slicing & 180 \\ \hline
12. & program analysis & 170 \\ \hline
13. & open source & 168 \\ \hline
14. & software development & 164 \\ \hline
15. & information retrieval & 153 \\ \hline
\end{tabular}
\end{table}

After inspecting the most used trigrams and 4-grams, we came to the conclusion that these phrases represent mainly clich\'es and bigrams complemented by insignificant words.

\section{Trends}

\subsection{Research Methods}

\begin{figure}[t]
\centering
\includegraphics[width=0.49\textwidth]{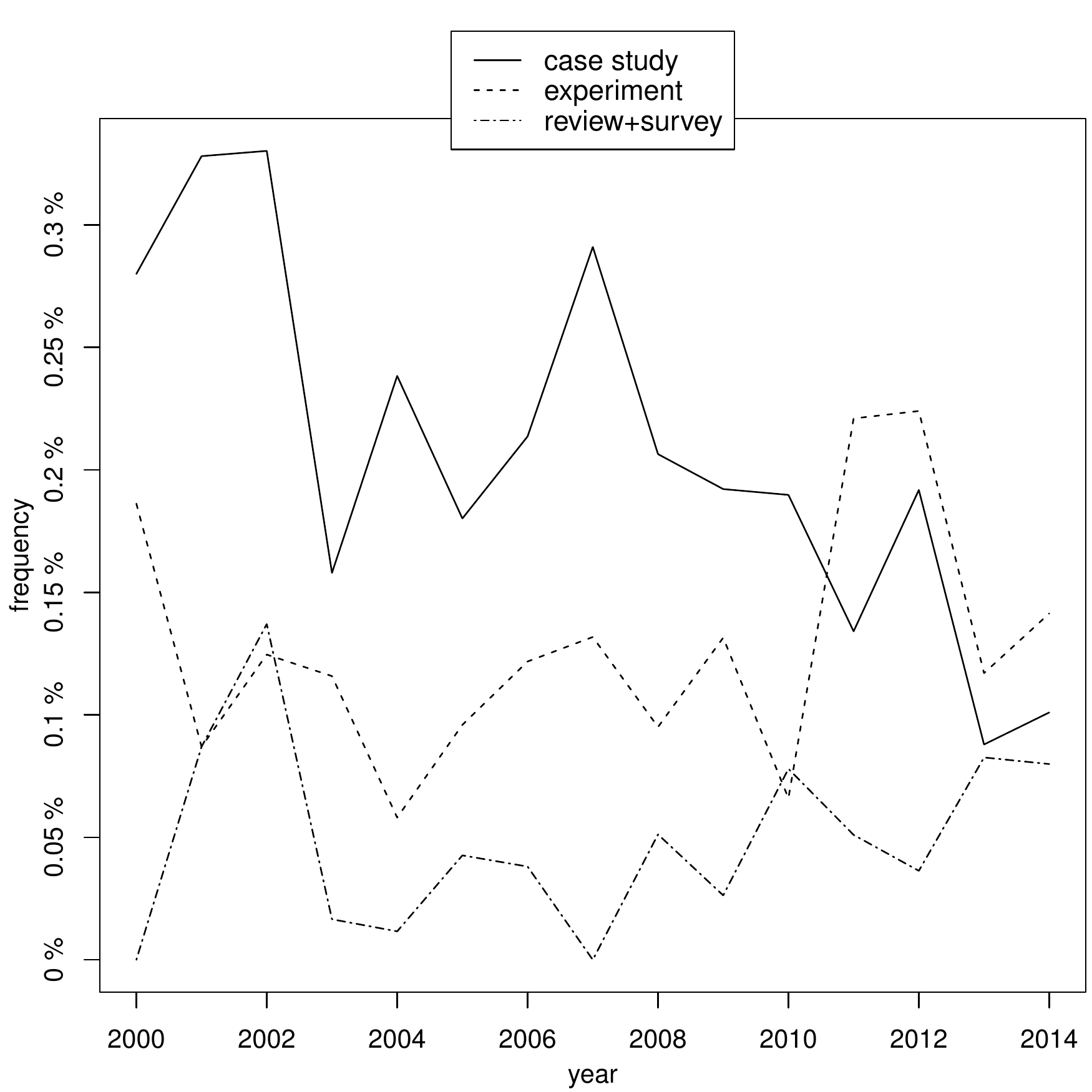}
\caption{Types of research in program comprehension.}
\label{fig:methods}
\end{figure}

We can see in Fig.~\ref{fig:methods} that case studies have a slightly decreasing tendency. There is a sudden rise of experiments in 2011, when they even outperformed case studies. Naturally, reviews and surveys are the least common types as they summarize existing research results.

\subsection{Techniques}

\begin{figure}[t!]
\centering
\includegraphics[width=0.49\textwidth]{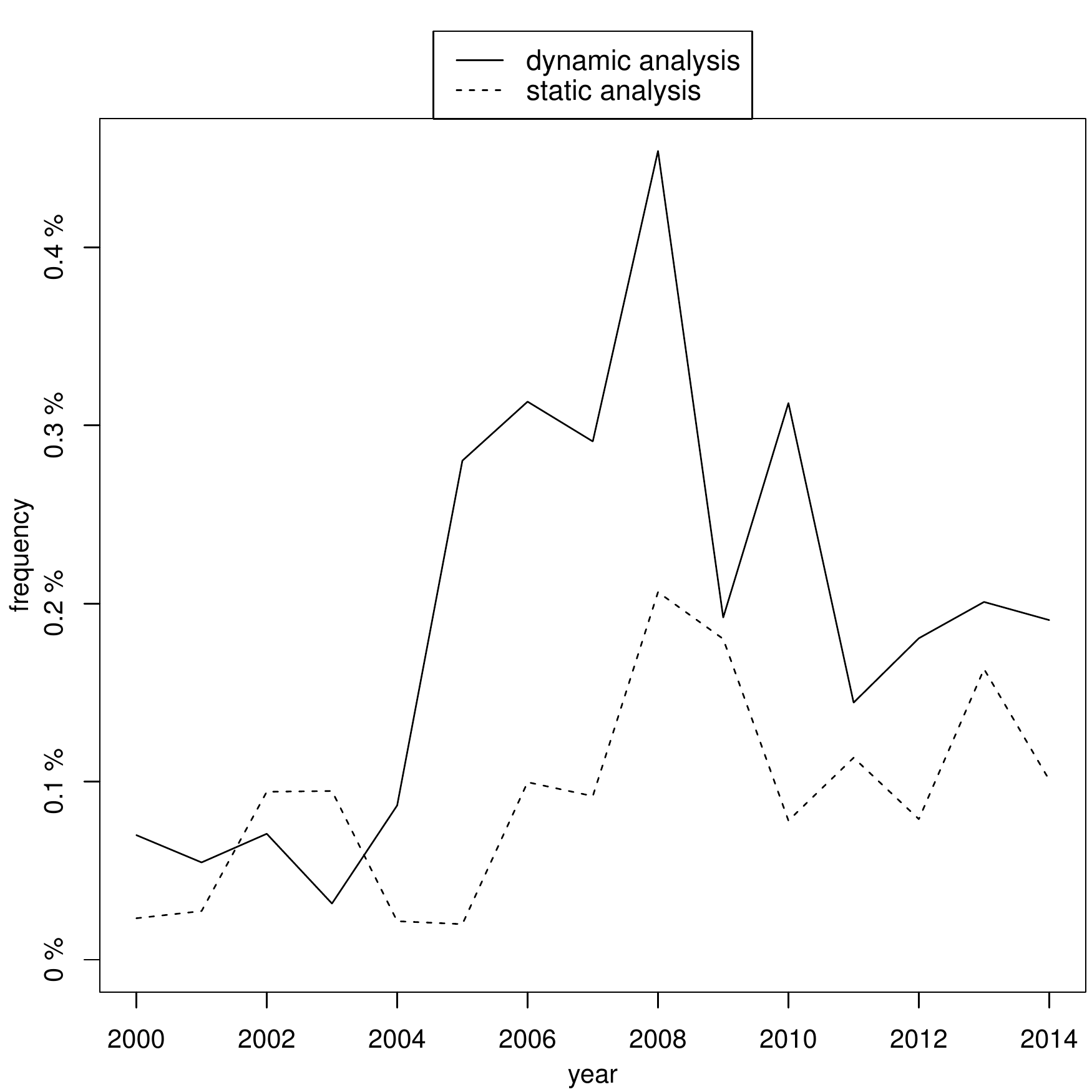}
\caption{Static vs. dynamic analysis.}
\label{fig:analysis}
\end{figure}

Static program analysis (e.g., \cite{Kollar11identification}) deals with the source code of an application without running it, whereas dynamic analysis (surveyed in \cite{Cornelissen09systematic} and exemplified in \cite{Bacikova13defining}) utilizes runtime information. As seen in Fig.~\ref{fig:analysis}, in 2004, the dynamic program analysis overtook the static one and this state lasts until today.

\begin{figure}[t!]
\centering
\includegraphics[width=0.49\textwidth]{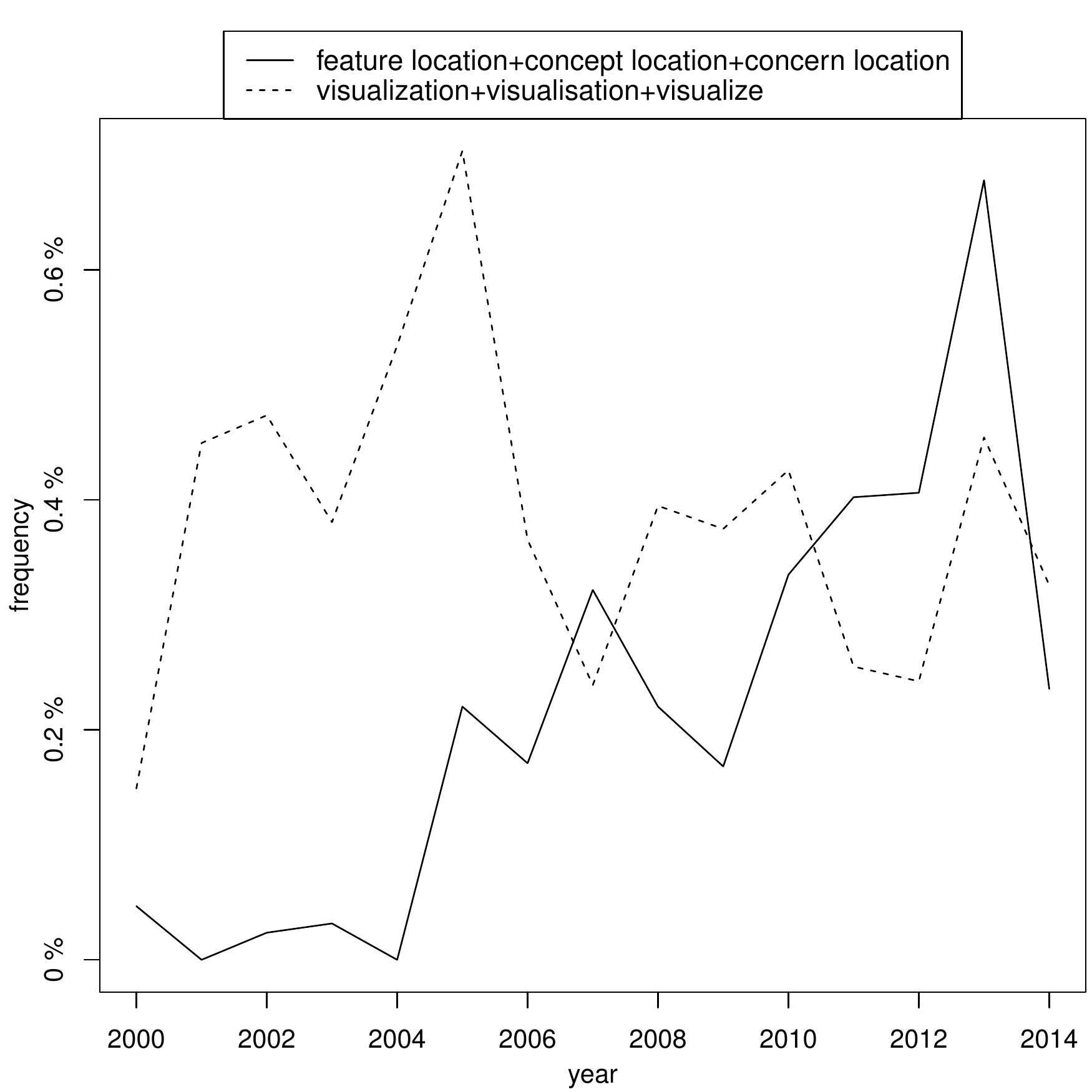}
\caption{Feature location vs. visualization.}
\label{fig:techniques}
\end{figure}

We can see a plot of two often used program comprehension techniques -- feature location and visualization -- in Fig.~\ref{fig:techniques}. While visualization is relatively steady, feature location rises rapidly from 2004.

\begin{figure}
\centering
\includegraphics[width=0.49\textwidth]{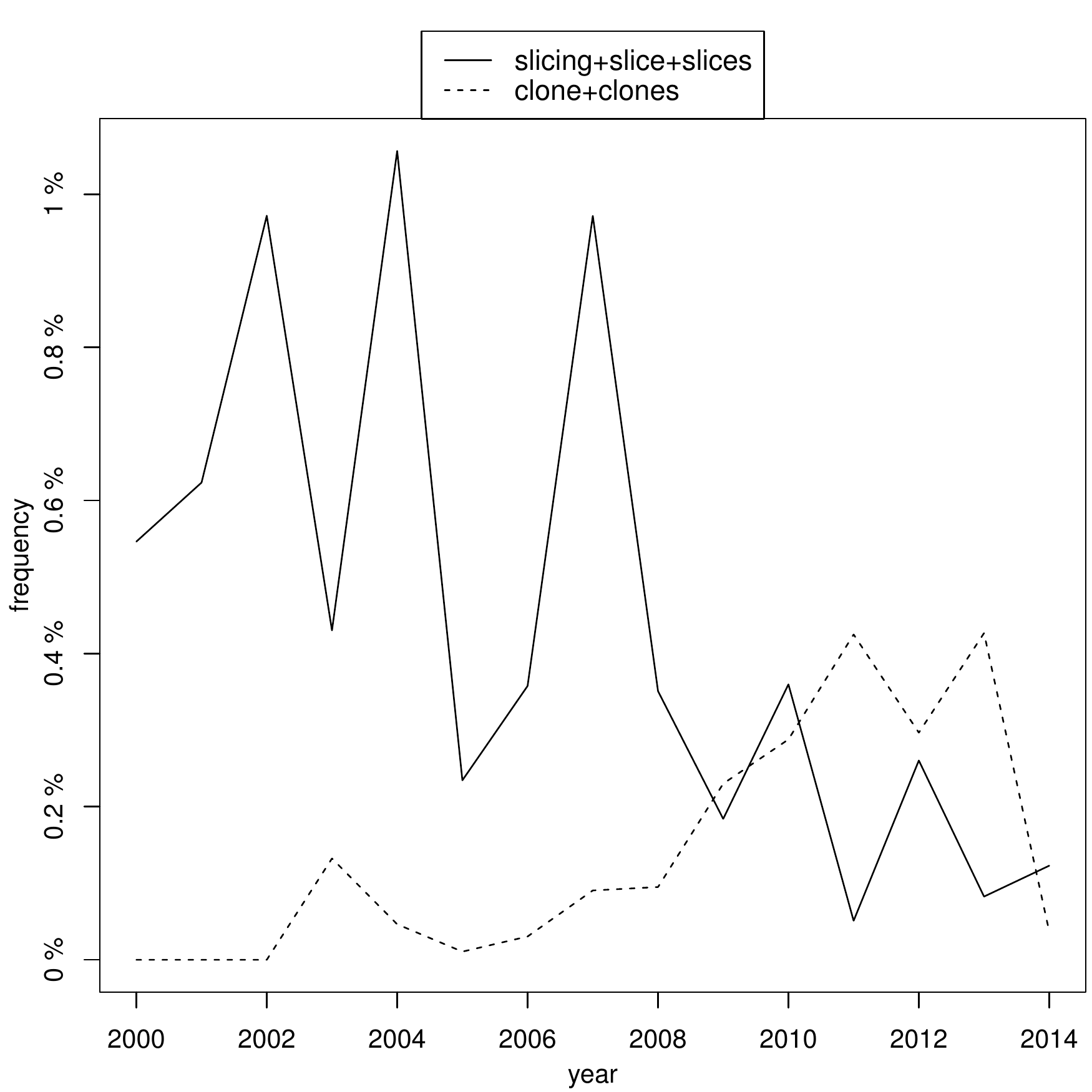}
\caption{Program slicing, code clone detection.}
\label{fig:slicing}
\end{figure}

Program slicing \cite{Xu05brief} is a technique used to find all code semantically related to the given statement or variable and produce a new program, containing only this related code. In Fig.~\ref{fig:slicing}, we can see a decreasing popularity of slicing.

Code clone detection \cite{Roy09comparison} is a research field with a long tradition. However, from Fig.~\ref{fig:slicing} it is obvious that it started to associate with program comprehension only in recent years.

\subsection{Systems}

\begin{figure}
\centering
\includegraphics[width=0.49\textwidth]{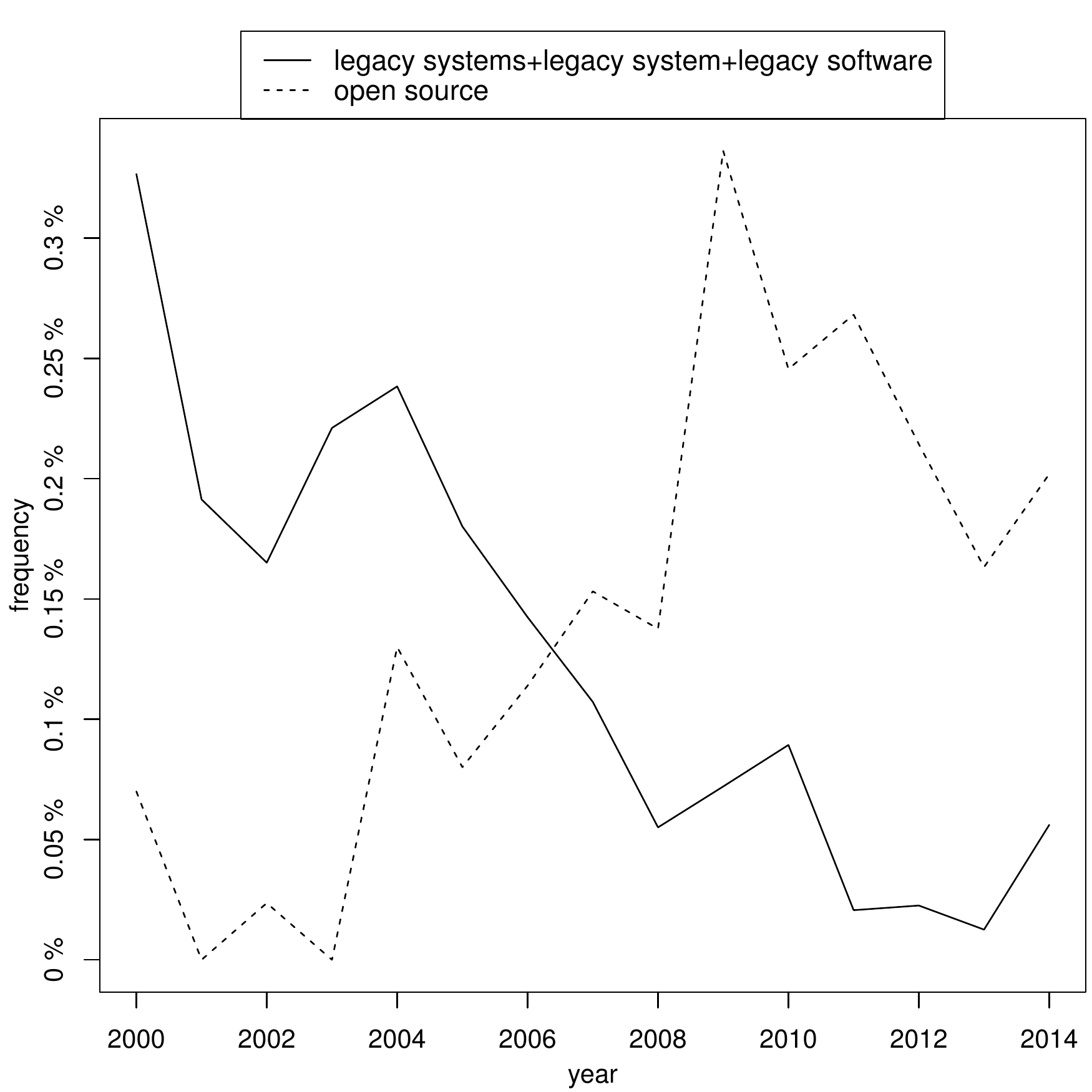}
\caption{Studied systems -- legacy vs. open source.}
\label{fig:subjects}
\end{figure}

In Fig.~\ref{fig:subjects}, we can see what types of systems the researchers study. Legacy systems have a clearly decreasing tendency during the last 15 years. One trend which immediately caught our attention is an extremely rapid increase of the n-gram ``open source'' between the years 2003 and 2009.

\section{Threats to Validity}

First of all, full texts of the articles were not analyzed. This is mainly due to the fact that many popular digital libraries forbid mass-downloading of the full texts. A notable exception, Scopus, has an option to download up to 50 articles at once. However, none of the few papers we tried to download were available in their full-text database. The most probable reason is the fact that publishers, which transferred the rights from the authors, often significantly limit publishing full-texts on other servers. A viable solution is open access publishing \cite{Brown03bethesda}.

A relatively small number of articles per year to produce reliable results is another possible threat.

\section{Related Work}

A similar trend analysis \cite{Demeyer13happy} was performed on the papers from nine Mining Software Repositories conferences. The authors analyzed another, albeit related, research field. Our study is methodologically inspired by this paper in some aspects, particularly the idea of n-gram frequency analysis. However, our approach differs in many ways. For example, the authors analyzed only nine years of one particular conference, while we performed a web search for the keywords. On the other hand, they analyzed full texts, while we only processed the metadata.

An n-gram query language syntax was inspired by Google Ngram Viewer \cite{Michel11quantitative}. We also considered directly using this tool for our study since it offers a text corpus many orders of magnitude larger than ours. While the time period is broader, too, it does not include the last two years. Furthermore, it is not specialized to computer science and the corpus consists of books, which are less suitable for latest trend analysis than research articles.

It is also possible to analyze Q\&A (question and answer) sites such as Stack Overflow to find out trending topics. In \cite{Barua14what}, a technique called latent Ditrichlet allocation \cite{Blei03latent} was used to recover topics from natural language texts.

\section{Conclusion and Future Work}

We analyzed the trending phrases in bibliographical references to see the rising and falling ones. To answer \textbf{RQ1}, the most rising trends are feature (and concept) location and the study of open source systems. Answering \textbf{RQ2}, program slicing and the study of legacy systems are the most falling trends.

Our next goal is to perform a manual systematic analysis \cite{Brereton07lessons} in selected subfields of program comprehension, using the acquired experience.

\bibliographystyle{IEEEtran}
\bibliography{../../Bibliography/bibliography}

\end{document}